# Spin-controlled Electron transport in Chiral Molecular Assemblies for Various Applications

Ritu Gupta,[a] Anujit Balo,[b] Rabia Garg,[c] Amit Kumar Mondal,[*c] Koyel Banerjee Ghosh,[*b] Prakash Chandra Mondal[*a]

[a]Department of Chemistry, Indian Institute of Technology Kanpur, Uttar Pradesh-208016, India

[b]Department of Chemistry, Indian Institute of Technology Hyderabad, Telangana-502285, India

[c]Institute of Nano Science and Technology, Knowledge City, Mohali, Punjab-140306, India

E-mail: amit@inst.ac.in (A.K.M.); koyel@chy.iith.ac.in (K.B.G.); pcmondal@iitk.ac.in (P.C.M.)

R.G. and A.B. equally contributed to this work.

**Abstract:**

The chirality-induced spin selectivity (CISS) effect has garnered significant interest in the field of molecular spintronics due to its potential for creating spin-polarized electrons without the need for a magnet. Recent studies devoted to CISS effects in various chiral materials demonstrate exciting prospects for spintronics, chiral recognition, and quantum information applications. Several experimental studies have confirmed the applicability of chiral molecules towards spin-filtering properties, influencing spin-polarized electron transport, and photoemission. Researchers aim to predict CISS phenomena and apply this concept to practical applications by compiling experimental results and enhancing understanding of the CISS effect. To expand the possibilities of spin manipulation and create new opportunities for spin-based technologies, researchers are diligently exploring different chiral organic and inorganic materials for probing the CISS effect. This ongoing research holds promise for developing novel spin-based technologies and advancing the understanding of the intricate relationship between chirality and electron's spin. This review showcases the remarkable CISS effect and its impact on spintronics, as well as its relevance in various other scientific areas.

**Introduction**

The study of spintronics is an intriguing area that delves into the impact of electron spins in various materials.[1–6] A particularly noteworthy development in spintronic research is the chirality-induced spin selectivity (CISS) effect, which has the potential to yield highly effective spin injection into metals or semiconductors through chiral materials, without the need for magnetic fields.[7–10] Alternatively, it can be stated that the CISS effect can facilitate the creation of spin-polarized electrons, without the need for a magnet, which could open up new possibilities for spintronics.[11–13] While most spintronic devices are made from (ferro)magnetic materials, the use of chiral organic molecules as spin filters, rather than traditional inorganic materials could lead to energy efficiency and device miniaturization. With a reliable source of spin-polarized electrons, ferromagnets can be magnetized through spin torque transfer. Organic-based spintronic devices are easy to fabricate and able to show the high spin selectivity that requires less magnetic field to switch the spin state from high to low resistance. Back in 1999, Ron and co-workers first observed the CISS effect using l- or d-stearoyl lysine-based chiral organic films.[14] This phenomenon revealed that photoelectron scattering asymmetry is over 100 times greater when electrons travel through chiral molecules. The CISS effect describes how chiral molecules can selectively transmit electrons with one spin over those with the other, meaning chiral molecules can act as spin selectivity. Essentially, electrons with specific spins can easily move through chiral molecules in one direction, depending on the handedness of the molecules.[15] While there is no complete quantitative explanation for the effect yet, it's thought that the electron moves in a helical electrostatic potential, which can help to explain this phenomenon. As an electron travels through the unique electrostatic potential of a chiral medium, it experiences an effective magnetic field that is perpendicular to its velocity. This magnetic field causes the energy levels of the electron's spin states to split, with one state gets stabilized and the other gets destabilized. The chiral structure creates a link between the electron's motion and spin, enhancing the transport of one spin over the other, depending on the handedness and direction of motion.[16] Moreover, this connection also impedes backscattering of the electron since reversing the electron's velocity direction necessitates a highly unlikely flip of its spin direction. As a result, the coupling between spin and velocity increases the competence of the electron transfer process through chiral materials.

Two primary methods for measuring the CISS effect are analyzing spin-dependent transport through chiral materials or examining the spin polarization induced by charge polarization in chiral materials.[11] To determine the CISS effect efficiency, spin polarization (SP) is calculated as $SP = \frac{(I\text{up} - I\text{down})}{(I\text{up} + I\text{down})}$, where $I_{up}$ and $I_{down}$ correspond to currents of spin-up and spin-down configurations.

This phenomenon has been verified in many chiral materials, DNA and oligopeptides, chiral metal-organic frameworks, amino acids, chiral supramolecules, polymers, proteins, and organic-inorganic hybrid materials.[11] The CISS effect bears relevance for various fields, ranging from spintronics, and enantioselective reactions to chiral recognition and quantum information. Effectively injecting spin has posed a challenge with such materials. Fortunately, the CISS effect in organic and inorganic molecules has provided a solution. These molecules serve as spin filters by displaying the CISS effect at room temperature, making it highly practical for spintronics applications. Researchers have been thoroughly exploring and refining various organic and inorganic materials to expand the limits of spin manipulation and create exciting prospects for spin-based technologies. Multiple experimental results have been compiled to gather information about CISS phenomena, enabling researchers to construct a knowledge base that predicts the CISS effect and applies it to various applications. This review proposes to deliver a broad summary of the numerous organic and inorganic materials utilized to investigate the CISS effect and the current comprehension of CISS-related phenomena.

**Chiral organic materials for spintronics**

Organic materials are attractive for spintronics because of their small spin-orbit coupling, which allows for long-range spin coherence over longer times and distances than conventional inorganic heavy materials. However, organic molecules typically could not be a good choice for spin injection because of their small spin-orbit coupling. The CISS effect is a promising development for chiral organic spintronics, as chiral molecules could transmit a preferred spin through a specific handedness when charge carriers are created.

Helicenes are molecules that contain fully conjugated carbon rings and lack stereogenic carbons. Due to the repulsion between their termini, helicenes adopt permanent helical conformations that can be either left-handed or right-handed. Kiran *et al.* reported that the enantiopure helicenes can be made oriented on a graphite surface and can act as efficient spin filters.[17] The authors have employed magnetic conductive probe atomic force microscopy (mc-AFM) in two distinct configurations to carry out the role of spin in the electron transport process through the molecular films. The results have significant consequences for understanding fundamental principles of spin-dependent transport in molecular systems, which have been further employed to design CISS-based electronic devices. The percentage of spin polarization (SP) was found to be + 49 % ± 3 % and – 45 % ± 3 % for *P* and *M* -enantiomer, respectively at ±1.0 V. In another study, Kettner *et al.* showed that helicenes with different strengths of spin-orbit coupling (SOC) when exhibit similar spin selectivity, suggesting the chirality of helicenes is instrumental in determining their spin-filtering capability.[18] Safari *et al.* demonstrated the first observation of the CISS effect through single chiral heptahelicene molecules

**(Figure 1)**.[19] The authors measured the CISS effect through heptahelicene molecules that are sublimed in vacuo onto ferromagnetic cobalt surfaces. The examination was carried out using spin-polarized scanning tunneling microscopy (SP-STM) at 5 K. The results revealed magneto chiral conductance asymmetries of up to 50% at 5 K. This study introduces a novel approach to analyzing single molecules, representing a significant advancement over prior studies in the field. The approach permits differentiation between the handedness of molecules with a given magnetization. This key discriminator provides a means to compare the conductance of enantiomers. The ability to compare enantiomer conductance in this manner is expected to provide valuable insights into molecular-level electronic properties and CISS mechanisms. Notably, the measurement was performed at a temperature of 5 K. The measured values are akin to those obtained at room temperature for cationic helicenes using the mc-AFM technique and a magnetic Ni-tip. The comparison of conductance asymmetry and polarization of different molecule-substrate presents a significant challenge for this approach. Suda *et al.* have reported a ground-breaking discovery related to the spin selectivity of chiral molecules that could be changed by external stimuli like temperature or light using an artificial molecular motor of overcrowded alkenes (OCAs) **(Figure 2)**.[20] A magnetoresistance (MR) device was fabricated which could validate the observation of spin-polarized current. Despite this, the device only yielded an insignificant MR value of approximately 2% due to unpolarized leakage currents. To better estimate the SP values, mc-AFM technique was utilized further. The thin films were made via spin-coated onto the HOPG substrate, followed by current-voltage (I-V) measurements using CoCr-based AFM tips **(Figure 2)**. The magnetization was oriented in either an upward or downward direction. At least, 40 I-V curves per magnetized direction were collected, and their average was obtained. Irrespective of the molecular chirality, there is always a higher absolute current under a positive bias. This is attributed to the work function variation between the HOPG substrate and the CoCr tip. The motor showed spin selectivity values of up to 44%.

Another study by Zhu *et al.* found first-generation molecular motors exhibiting multistate switching of spin selectivity.[21] There are four chiral states in which these compounds exist, which can be accessed through external stimuli like light and heat. Raman spectroscopy reveals that the molecular motors undergo isomerization when deposited on gold substrates, triggered by exposure to light and heat. This exceptional system could be interconverted in a specific arrangement, either clockwise or counterclockwise. The mc-AFM measurements confirmed the ability of molecular motors to act as spin filters. The irradiated samples underwent a successful conversion, predominantly yielding diastereoisomers despite a moderate proportion of metastable to stable isomers. This highlights the significant importance of the spin-polarization in the mentioned studies. The study also showed that the right-handed motor in the photostationary state exhibits efficient spin polarization and the modulation of spin filtering properties by regulating the sequence of helical states.

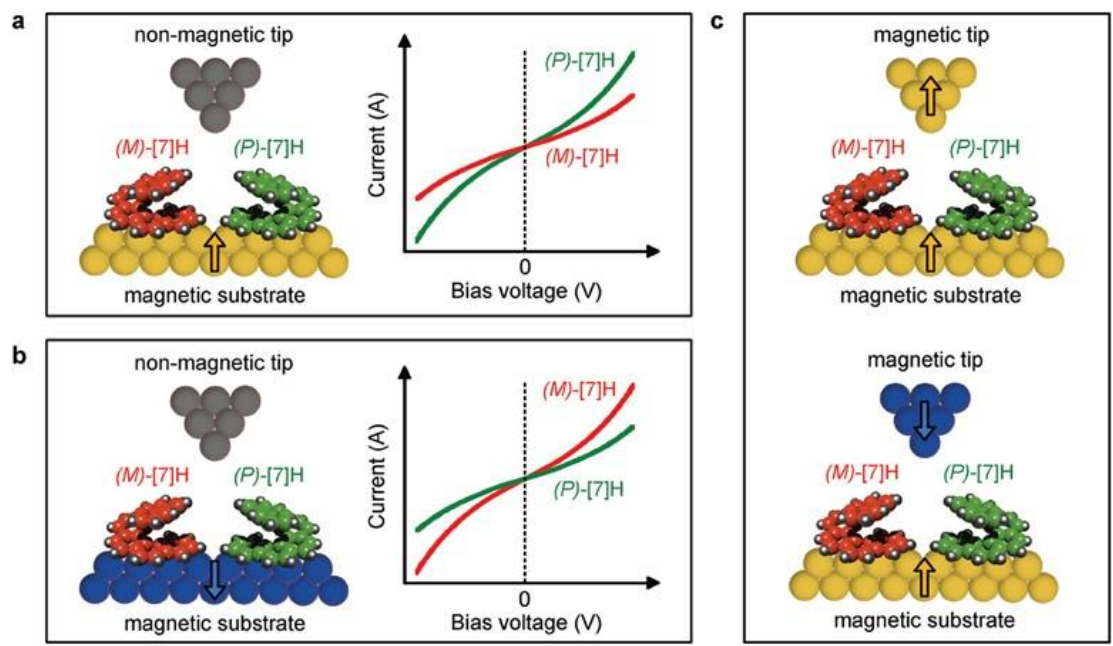

**Figure 1**. (a-b) Demonstrating the magneto-chiral electron tunneling on the ferromagnetic electrode for heptahelicene enantiomers. (c) Measurements with a magnetic STM tip on a ferromagnetic substrate. Figures are adapted with permission from ref.[19] Copyright 2023 Wiley.

Other than this, various biomolecules, including DNA, oligopeptides, and proteins such as cytochromes and bacteriorhodopsin, have been incorporated for the CISS effect studies due to their preferred handedness.[22,23] Recently, Gupta *et al.* fabricated a soft biomolecular junction by a flip-chip approach, where Ustilago maydis Rvb2 protein, an ATP-dependent DNA helicase, was sandwiched between two conducting substrates: Ferromagnetic Ni and indium tin oxide (ITO).[24] In the present study, the authors utilized novel Ustilago maydis Rvb2 protein; their hexameric and helical structures and room temperature stability made it competent for the CISS studies. The Ni substrates were covalently modified with a Glutathione (GSH) linker to electrostatically bind the Rvb2 protein, and the junction was completed by using the PMMA gel-coated ITO electrode as the top electrode, as shown in **Figure 3a**. Interestingly, all devices were fabricated without using any expensive cleanroom facility used for vacuum top-electrode deposition. The vacuum-deposited top electrode on the protein film can often denature the protein film and, hence, can reduce the device's performance. Thus, the current work illustrates a versatile and facile approach to creating a stable mechanical contact underneath the Rvb2 protein film.

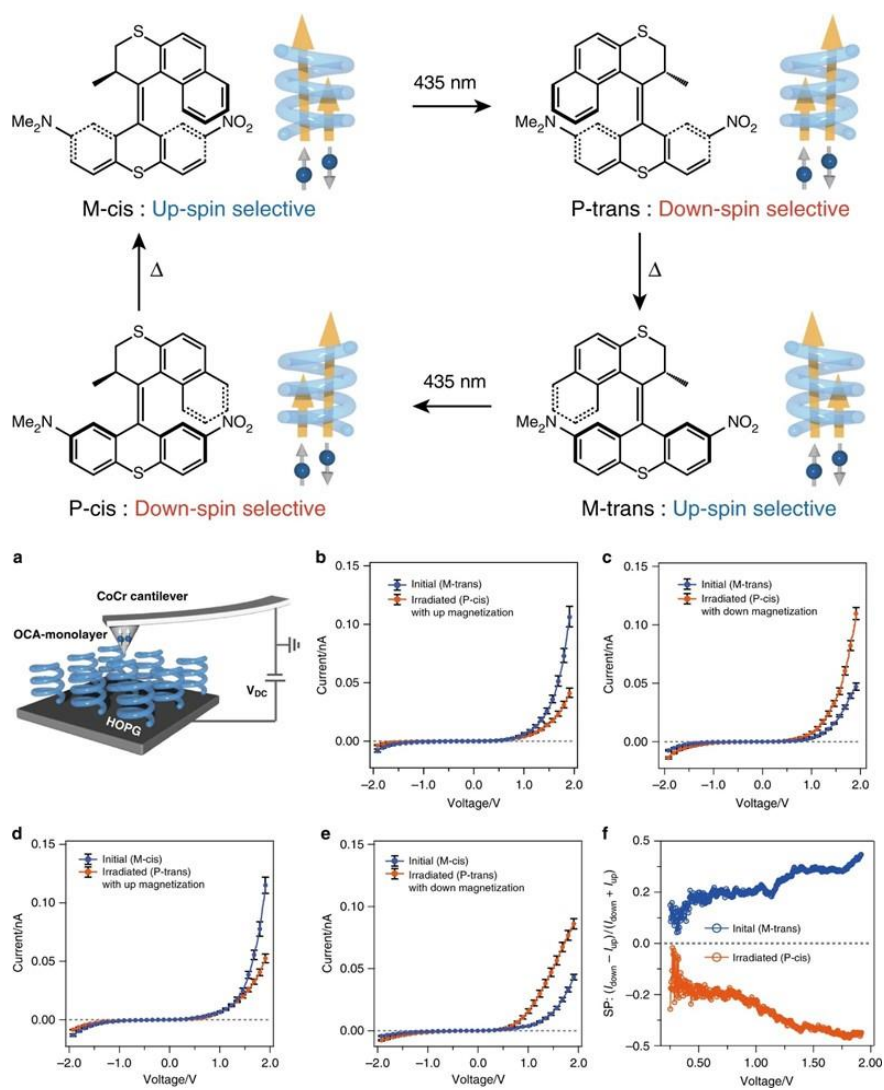

**Figure 2**. Molecular structures and rotation cycles of an artificial molecular motor of overcrowded alkenes. (a) Schematic representation of mc-AFM measurements. Representations of I–V plots for M-trans isomers before and after irradiation with tip magnetized with up (b) and down (c) magnetic field orientations. Representations of I–V plots for M-cis isomers before and after irradiation with tip magnetized with up (d) and down (e) magnetic field orientations. (f) Plot of spin polarization vs applied voltage before and after the irradiation. Figures are adapted with permission from ref.[20] Copyright 2019 Nature.

In comparison to the DOWN-magnet and no-magnet conditions, the biomolecular junction demonstrated higher current density and differential conductance in the UP-magnet condition, an evident signature of the CISS effect (**Figure 3c-3d**). More importantly, the Rvb2 junction illustrated a 30 % spin-polarization at 500 mV bias at room temperature, even at a thicker film of ~ 266 ± 12 nm (**Figure 3e**). Both the asymmetric density of states of Ni substrates at the Fermi energy level and the spin-filter nature of Rvb2 protein due to their preferred handedness contributed to the long-range spin-selective charge transport in the biomolecular junction (**Figure 4**). The covalent interface between the ferromagnetic Ni substrate and GSH further assisted in enhancing the spin diffusion length without spin decoherence. Thus, the overall work supports the room temperature operation of a spintronic device, which is a prerequisite for commercialization.

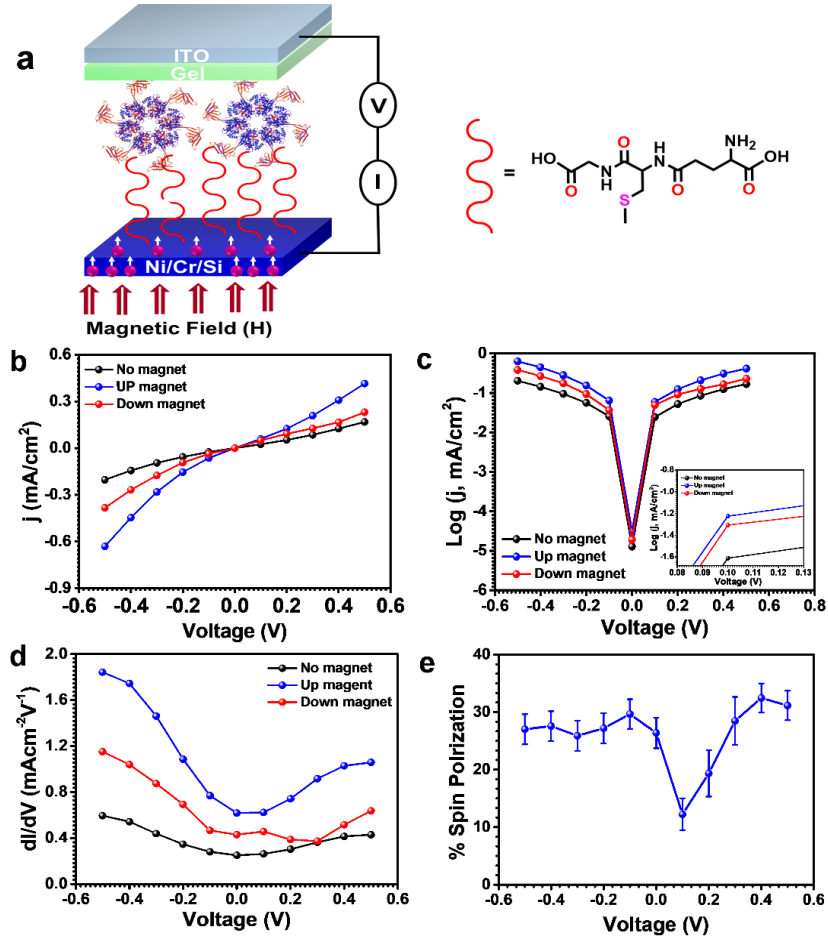

**Figure 3.** Spin-selective charge transport in Rvb2-based biomolecular junction. (a) Schematic illustration of the Ni/Rvb2/ITO biomolecular junction. (b) j-V curves (c) semilog curves and (d) derivative conductance curves of Ni/Rvb2/ITO biomolecular junction in the absence and presence of applied magnetic field. (e) Spin polarization as a function of applied bias. Figures are adapted with permission from ref.[24] Copyright 2023 American Institute of Physics.

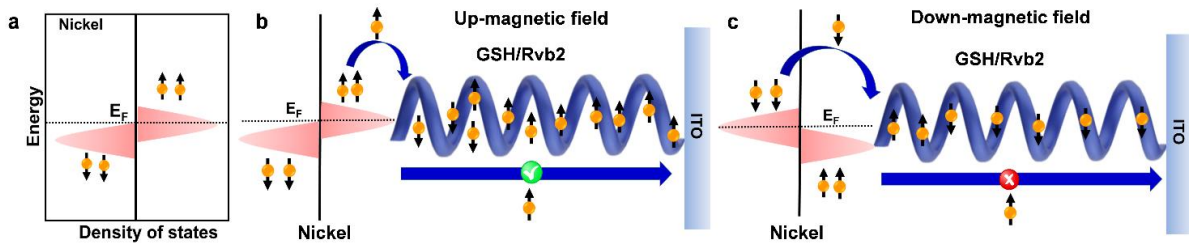

**Figure 4.** Mechanistic model for spin-selective charge transport in Rvb2-based junctions. (a) Energy vs. density of states plot for Ferromagnetic Ni substrate in the absence of applied magnetic field. (b-c) Proposed spin-selective charge transport in Ni/Rvb2/ITO junction in spin-UP and spin-DOWN magnetic field conditions. Figures are adapted with permission from ref.[24] Copyright 2023 American Institute of Physics.

Chirality plays a vital role in spin filtering in various organic polymers and π-conjugated molecules. The chiral polymers and supramolecular structures are a great alternative for high spin selectivity due to the CISS effect. For example, Mondal *et al.* reported that cysteine-thiophene-based polymers exhibited high spin filtering properties.[25] It has been shown that the critical significance of spin-selective transport in supramolecular structures.[26] This was accomplished using a helical chiral structure, which facilitated the intended process through the CISS effect. A recent study that was conducted by Kulkarni and co-workers using a helical supramolecular outweighs the presence of chiral centers in a supramolecular structure.[27] To illustrate this finding, researchers utilized coronene bisimide-based organic molecules with chiral alkoxyphenyl side chains to form nanofiber structures. These fibers were transferred onto a magnetized nickel-gold surface and the magnetization was placed perpendicular to the surface plane. The AFM tip was set at ground potential and a different bias was applied to the substrate. Nanofibers are bundled due to π-stacked aggregation and the height was calculated to be 17 ± 3 nm by AFM analysis **(Figure 5)**.

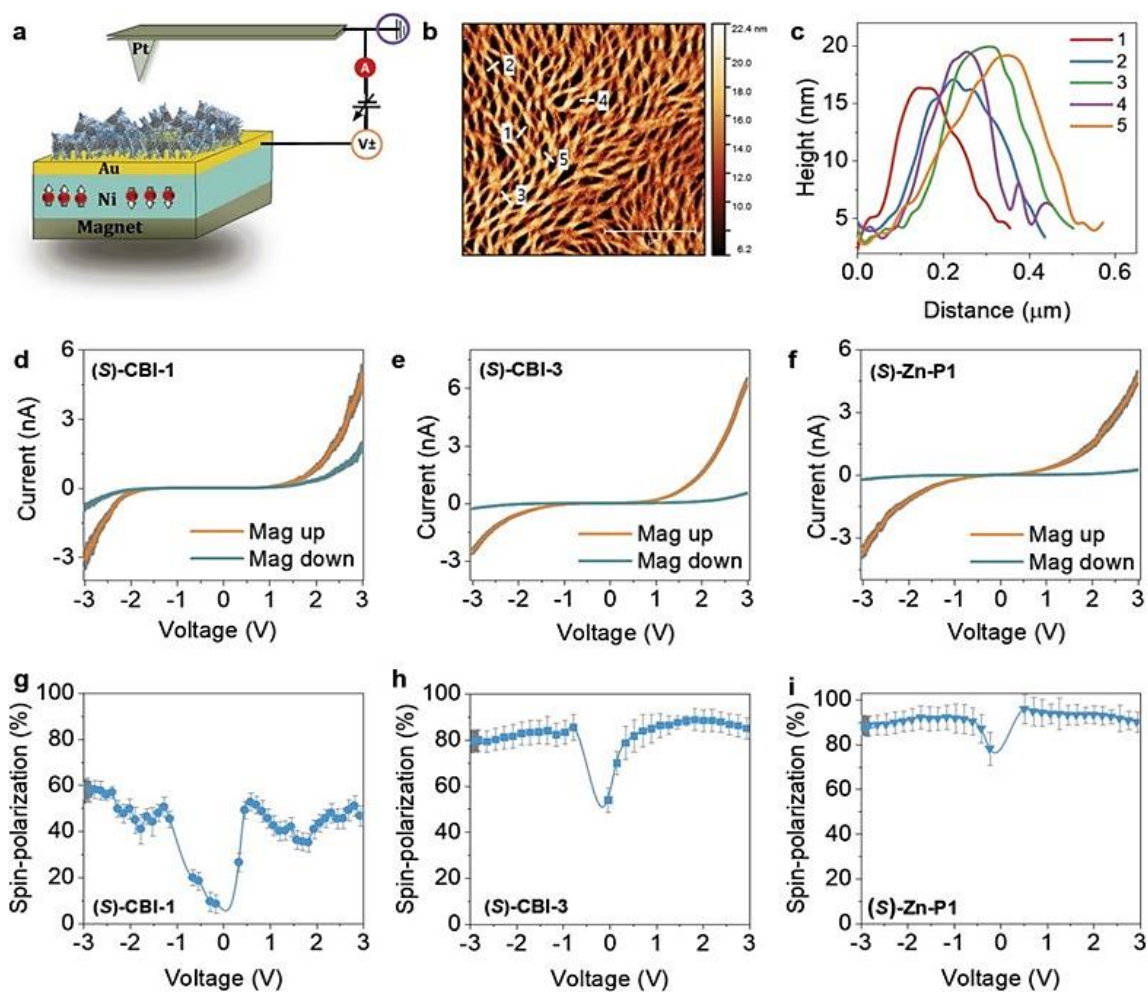

**Figure 5**. (a) Representation of mc-AFM setup. (b-c) AFM image of the supramolecular nanofibers and AFM height profiles. (d–f) The averaged I–V plots for different samples where Ni film magnetized with the north pole pointing up (orange) or down (cyan). (g–i) The plot of SP% vs voltage for different samples. Figures are adapted with permission from ref.[27] Copyright 2020 Wiley.

The spin filtering efficiency was consistently more than 80% at room temperature with thicknesses ranging from 10 to 40 nm. Furthermore, the weak dependence on thickness suggests that obtaining high polarization values for thinner nanofibers is an effective means of ensuring the maintenance of these values even for thicker ones. Based on these findings, it could be established that the magnitude of spin polarization (SP) in this range of heights is reliably high and persistent, which has important implications for future research in this area. It was also demonstrated that SP values could be further manipulated by mixing the chiral and achiral assemblies through the chiral amplification effect. The authors have reported a strong correlation between the molecular structure and the spin-filtering efficiency **(Figure 6)**. Typically, the supramolecular helicity would change based on the stereo configuration of the chiral centers in the molecules, however, the study showed that the same L-chirality of the molecule can also produce left and right-handed helicity and that the spin filtering property could be inverted by changing the temperature **(Figure 6)**.

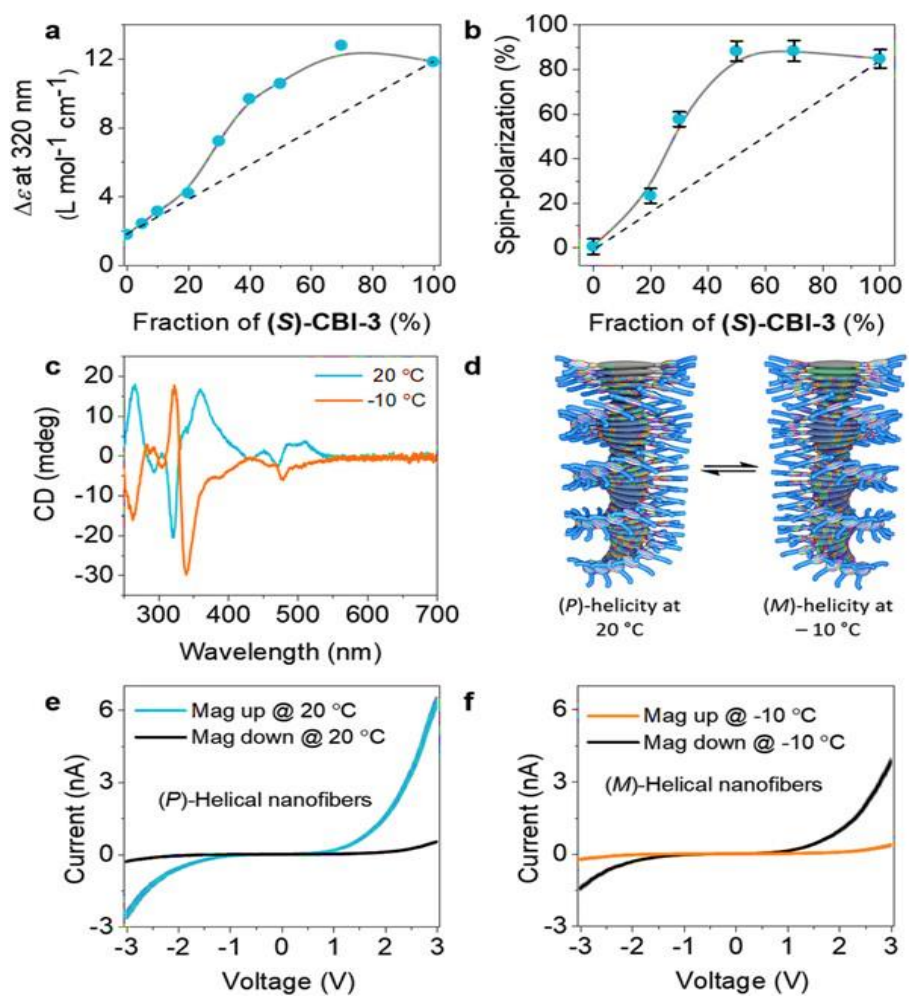

**Figure 6**. (a) Plot of molar circular dichroism (Δε) vs fraction of chiral molecules. (b) Spin polarization values at +3 V for different fractions of samples measured by mc-AFM. (c) Temperature-dependent CD plots. (d) Schematic representation of helical chirality depending on different temperatures. e,f) The averaged I–V plots for chiral samples at different temperatures. Figures are adapted with permission from ref.[27] Copyright 2020 Wiley.

Previous studies for understanding the CISS effect at the molecular level, however, did not consider the impact of the polymer backbone on the spin-filtering property. To address this issue, Mishra *et al.* reported the CISS effect where the spin-polarized electrons transferred to the primary molecular axis of the polymer backbone.[28] The authors used thiol (-SH) functionalized poly(phenylacetylene)s that were adsorbed as self-assembled monolayers on the gold surface. The spin polarization was approximately 53 and 56%, resulting in a spin transmission ratio of approximately 1:3 **(Figure 7)**. Interestingly, there was no discernible difference in the averaged I-V plots for optically inactive polymer samples in the presence of different magnetic field orientations. Earlier CISS-based mc-AFM measurements indicated a non-linear current-voltage relationship. The two spins exhibited a distinct threshold for charge injection, indicating no spin-flipping process involved. The spin injected at a lesser potential depends on the chirality of polymer samples, with a difference in injection threshold of roughly 100:10 meV **(Figure 7)**. For D-polymer, the lower spin corresponds to the magnet pointing down, which means that the injected spins are polarized parallel to the electron's velocity. To investigate the feasibility of these polymers for spintronics applications, the spin valve device was fabricated, and the magnetoresistance (MR) effect was further measured. The device's resistance was accurately determined using four probe electrical contacts. Unlike other MR devices, the MR behavior of this device exhibited asymmetry relative to the field sign **(Figure 7)**, which is attributed to the use of one ferromagnetic electrode. The MR values are relatively low likely due to the pinholes in the monolayers and electron scattering.

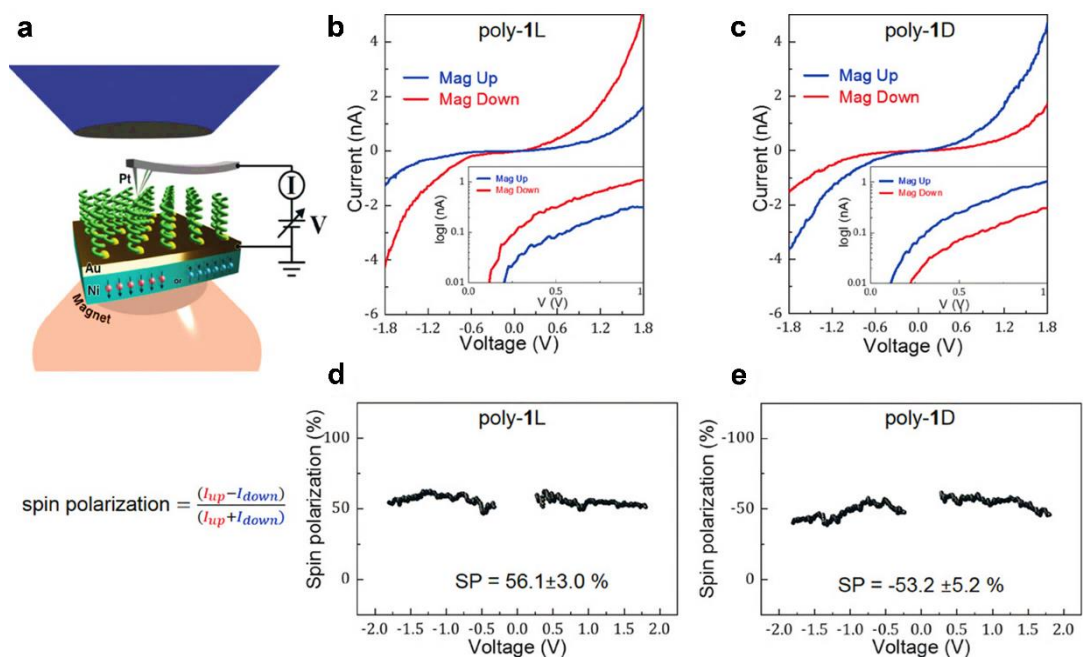

**Figure 7**. (a) Scheme of mc-AFM setup. (b-c) Averaged I–V plots for L and D polymers, respectively. (c-d) The spin polarization vs voltage plots for L and D polymers, respectively. Figures are adapted with permission from ref.[28] Copyright 2020 Wiley.

Previous studies have observed spin-selective transport in supramolecular structures that use homochiral building blocks with stereocenters. The question can be asked if the presence of stereocenters is crucial to observe the CISS effect. This is a crucial point for consideration, as it could greatly impact our understanding of this phenomenon. A recent report by Mondal *et al.* has demonstrated that the spin selectivity is not due to individual stereocenters of supramolecular structures but rather originates from helical supramolecular nanofibers.[29] The authors successfully demonstrated that the CISS effect originates from supramolecular structures having only achiral monomers. They have employed triphenylene-2,4,10-tricarboxamide motifs that possess helical chirality to investigate the spin polarization of the samples **(Figure 8)**. The helical chirality was deliberately manipulated to selectively produce opposite helices through the application of chiral solvents. The results indicate that the SP values increased with the sample's thickness, and the effect was found to be non-local, according to the intensity of the CD signals of the polymers. The thickness dependence can be attributed to at least two mechanisms operating namely the CISS effect and the scattering. Moreover, the temperature dependence supports that phonons may contribute to the CISS effect.

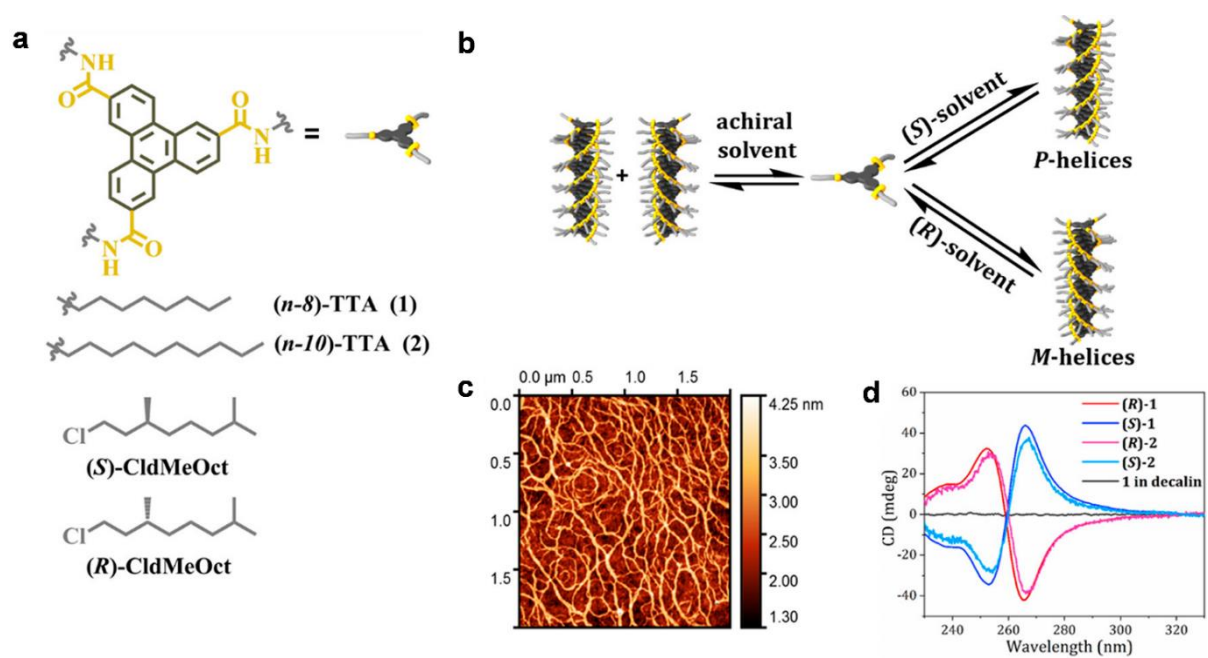

**Figure 8**. (a) Chemical structures of achiral molecules and chiral solvents. (b) Representation of chiral self-assembly processes by solvent molecules. (c) AFM image of the nanofibers. (d) CD spectra of molecules made from chiral (*S*)-and (*R*)-solvents. Figures are adapted with permission from ref.[29] Copyright 2021 American Chemical Society.

The recent work by Rösch *et al.* demonstrated a significant contribution to the CISS effect using helical order in the quarine dyes, rather than the chiral centers present in the dyes.[30] The efficient spin filtering ability is mainly due to the chiral molecules that were typically organized through the anchoring groups or supramolecular polymers. Recently, Labella *et al.* reported a new type of spin-filtering material based on bowl-shaped subphthalocyanine-based molecules, and the spin polarization was nearly 50%.[31] From the I-V characteristics, it is clear that the magnetization direction of the substrate is a crucial factor in determining the current exhibited by both the enantiomers. Precisely, the *M*-enantiomer shows higher currents when the substrate is magnetized with the magnetic field up. Conversely, the current is lower in the case of *P*-enantiomer when the same magnetization is applied. Notably, the spin polarization values are relatively high, despite the absence of any anchoring groups or supramolecular polymerization. This underscores the remarkable self-organizing ability of subphthalocyanines in thin films. These promising findings suggest that bowl-shaped aromatic molecules could offer practical and effective materials for CISS-based spintronic devices.

Giaconi *et al.* have recently demonstrated the high spin selectivity of above 60% in self-assembled monolayers of thia[4]heterohelicenes prepared on a gold electrode.[32] From mc-AFM measurements, it has been found that spin-selective electron transport occurred at low voltages of ± 0.3V. Zhang *et al.* very recently reported topologically chiral knot molecules without having no presence of stereogenic atoms still behave as a new type of spin-filtering material at room temperature **(Figure 9)**.[33] The intrinsic chirality has been confirmed by CD spectra and correlated with very high spin polarization of up to 88% as evident from mc-AFM measurements. The Λ trefoil knot exhibited higher current levels when the tip was magnetized upwards, while the Δ trefoil knot displayed the opposite trend, with higher current recorded when the tip was magnetized downwards.

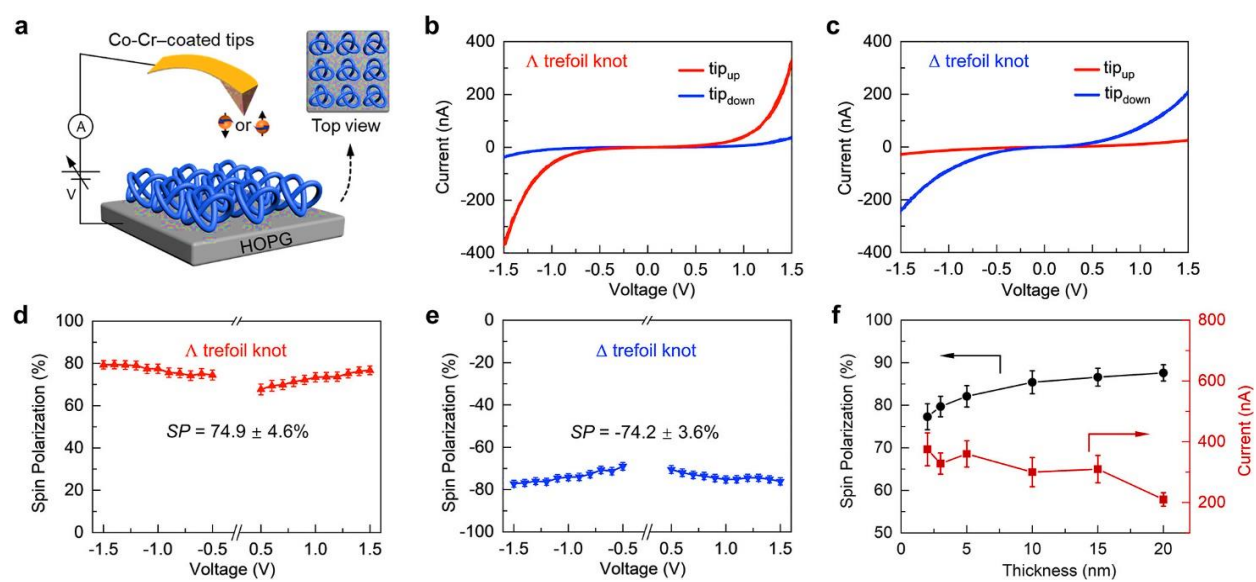

**Figure 9**. (a) Representation of mc-AFM setup. The averaged I-V plots for Λ (b) and Δ (c) molecular trefoil knots. Plot of spin polarization vs applied voltages for Λ (d) and Δ (e) knots. (f) Thickness-dependent SP% values and current intensities. Figures are adapted with permission from ref.[33] Copyright 2023 American Chemical Society.

In a controlled experiment, the samples containing a 50/50 mixture of ∧ and Δ trefoil knots were tested, and there was no discernible difference under two different magnetic field directions. The spin polarizations for the ∧ and Δ molecular trefoil knots were calculated to be 74.9 ± 4.6% and -74.2 ± 3.6%, respectively. Accordingly, the chiral knots provide a viable opportunity for the advancement of a novel class of spintronics elements. This innovative approach holds significant promise for evolving the field of organic spintronics, given the unique properties of chiral knot molecules.

**Organic-inorganic hybrid materials for spintronics**

The employment of inorganic-organic hybrid materials in various relevant applications reveals its potential as a promising material as a spin filter. It is demonstrated that the functionalization of inorganic material with chiral organic molecules results in spin-polarization in the hybrid materials at ambient temperature. Besides, utilizing various types of organic molecules, long-range spin-selective electron conduction through inorganic-organic hybrid materials is possible. To discuss the exploration of different chiral inorganic-organic hybrid materials in detail, this section is divided into the following sub-sections.

**Magnetoresistance devices**

The importance of inorganic-organic hybrid materials in the fields of photovoltaic,[34] optoelectronic, and electrochemical devices are well documented. In recent years, its relevant application in the field of spintronics has also been proven. There are many reports demonstrating the conduction of the spin-selective electron through the organic-inorganic hybrid system using ferromagnetic (FM) materials.[35] It has been found that a spin-valve consisting of two colossal magnetoresistance (CMR) half metallic electrodes and an organic semiconductor **(Figure 10)** situated in between two, resulted in large magnetoresistance for the channel length of the device lower than 140 nm whereas, the magnetoresistance is absent for larger channel length. This observation supports that spin-polarized transport occurs through the organic semiconductor. Carrier injection probability for lower channel length is proportional to the acceptance probability by the second ferromagnetic electrode. Thus, the resistance for parallelly oriented electrodes with the same spin alignment decreases and is responsible for the magnetoresistance signal. The spin polarisation is missing for longer channel lengths i.e., for longer organic semiconductors.

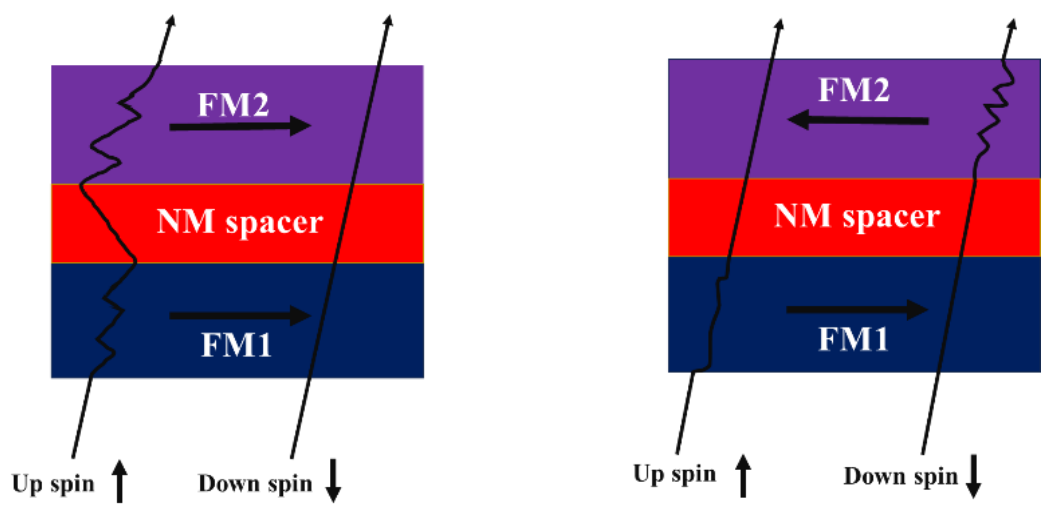

**Figure 10.** Schematic representation of a spin-valve structure demonstrating spin-dependent scattering at the two ferromagnetic electrodes in parallel and antiparallel alignments as shown in the literature. The figure is reconstructed from ref.[35]

Spin injection into the organic material and coherent spin transport through it, are considered two processes of spin polarization effect.[36] Significant enhancement of the spin transport efficiency has been reported by employing an *n*-type semiconducting polymer P(NDI2OD-T2) as the nonmagnetic spacer in spin valves by Li *et al.* It deciphers that an exceptionally high magnetoresistance can be achieved at room temperature by improving the semiconducting polymer and Co FM electrode interface by introducing a thin layer of $AlO_x$ as well as optimal annealing of bottom $La_{2/3}Sr_{1/3}MnO_3$ (LSMO) electrode. Further, it has also been found that the transportation through the polymeric interlayer is spin-dependent and the spin diffusion length is weakly dependent on the temperature.[37] After the invention of the CISS effect, numerous organic-inorganic hybrid materials have been explored for spin-selective electron transfer. As demonstrated by Bloom and others, the CdSe quantum dots functionalized with chiral cysteine molecule allows spin-selective electron transfer **(Figure 11)**, which were investigated using magnetic conductive-probe atomic force microscopy (mCP-AFM) measurements as well as magnetoresistance measurements.[38] It is also shown that chirality-modified metal-organic crystals (MOCs) e.g., Cu(II) or Co(II) phenylalanine enable long-range, spin-selective electron conduction at ambient temperature.[39,40] The metal ions are organized in two-dimensional layers between either L- or D-enantiomers of pentaflurophenylalanine ($F_5$Phe) or phenylalanine (Phe).

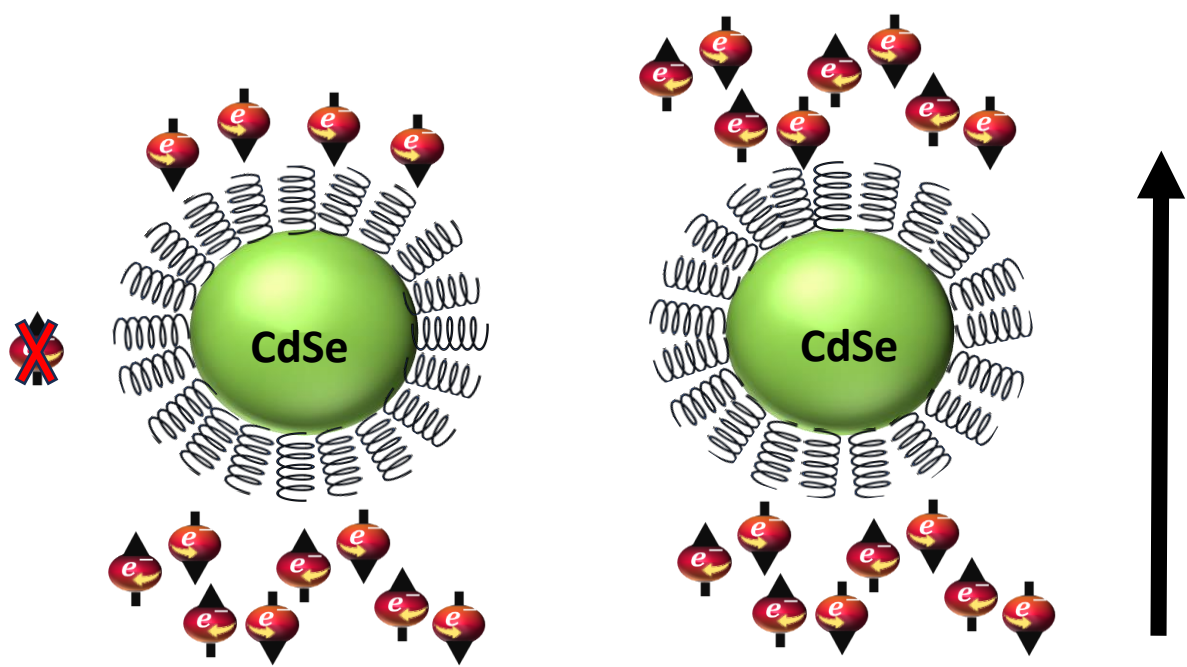

**Figure 11.** Schematic representation illustrating charge conduction via cysteine functionalized CdSe quantum dots with spin selectivity.

The spin-selective electron conduction through chiral enantiomers studies performed by mCP-AFM has revealed higher current for L-enantiomers of phenylalanine with up spin magnetization and opposite behavior has been observed for D- enantiomers **(Figure 12)**. It has been proposed that the preferred spin injection depends on handedness (L or D) and preferred spin transport depends on the magnetization direction of the molecular ferromagnet, which is independent of chirality. So, the different current magnitudes for L- and D- enantiomers with up and down spin magnetized substrates have been observed. The unusual observed thermally activated ferromagnetic behavior of this Cu-phenylalanine crystal can be explained by an indirect interaction between the unpaired electrons on the Cu (II) ion, that is mediated through the chiral lattice, forming a low-lying thermally populated ferromagnetic state while for Co-phenylalanine crystal due to slight change in crystal structure, it is showing antiferromagnetic interaction. These studies truly engross the combining effect of paramagnetic ions and chiral lattice and expanded to the organic molecule-based spintronics applications.

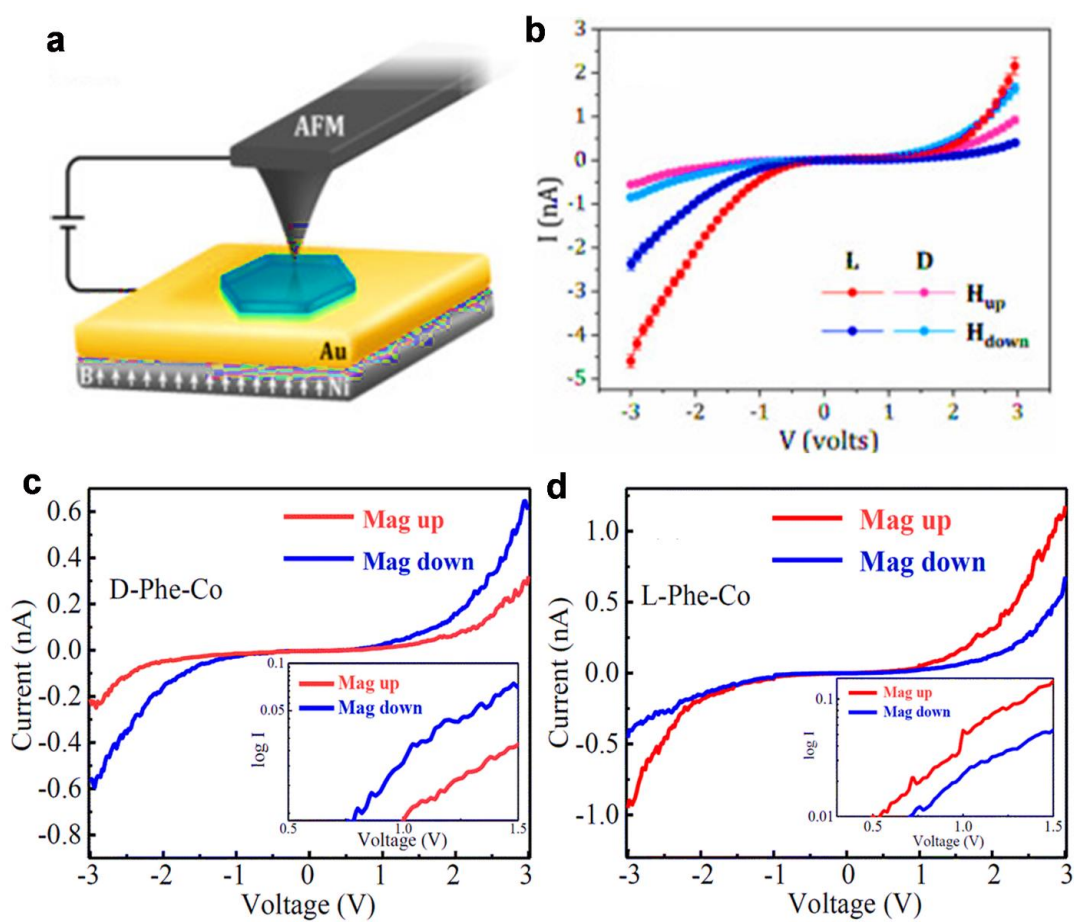

**Figure 12.** Spin-selective conduction. (a) Schematic representation of the mCP-AFM measurement. The catalyst is coated on gold coated Ni surface and a conducting AFM tip is in contact from above. An external magnetic field is used to magnetize the substrate. The current-voltage (I-V) measurements plots of (b) D- and L-Phe-Cu crystals magnetizing the substrate with different magnetic orientations. (c, d) The current-voltage (I-V) measurements plots of D- and L-Phe-Co crystals measured at room temperature. The log I-V plot is shown in the insets in (c) and (d). Figures a and b are adapted with permission from ref.[39] and figures c and d are adapted with permission from ref.[40]

**Photovoltaic & light emitting diode**

The spin-filtering property of chiral molecules offers great opportunities in photoinduced spintronics applications. After the invention of the CISS effect, several studies demonstrate chiral molecular assemblies exhibiting spin-polarization at room temperature. Spin polarization has been achieved by utilizing photoinduced charge separation and injection from well-coupled organic chromophores or quantum dots like CdSe at room temperature.[41] Exploiting CISS, the organic bulk heterojunction's photoconversion efficiencies can be improved by spin-selective charge transport. Zhang and others explored the reduction of the recombination of polaron pairs at the interface of poly(3-hexylthiophene/1-(3-(methoxycarbomyl)propyl)-1-1-phenyl)(6,6)$C_{61}$ (P3HT-PCBM) can be achieved by doping with spin-polarized galvinoxyl radical which results in increasing the efficiency by 18%.[42] They proposed a spin-flipping mechanism that suggests an exchange interaction between spin-polarized radicals and charge acceptors resulting in the suppression of the polaron pairs recombination rate by converting their spin from singlet to triplet state **(Figure 13a)**. Giovanni and others highlighted low-temperature solution-processed organic-inorganic hybrid halide perovskite $CH_3NH_3PbI$, as a promising candidate for its ultrafast spin-switching capability in photovoltaics and light-emitting devices, irrespective of their large spin-orbit coupling and small spin electron-hole relaxation lifetime.[43–45] It has been observed that chiral semiconductors-based quantum dot thin films can transmit a particular spin-oriented electron preferentially. Recently Das et al. reported chiral (R-/S-MBA)$_2$CuCl$_{4-x}$Br$_x$ compounds with tunable band gaps for photovoltaic and photocatalytic applications.[46]

It has been revealed that the lack of spin-polarized charge carriers in semiconductors at room temperature limits its application in opto-spintronics. A hybrid inorganic/organic spin light-emitting diode (h-OLED) has also been found to produce ~80% giant magneto-electroluminescence with the emission of red, green, and blue lights.[47] Recently it has been demonstrated that spin-polarized charge carriers can be injected into hybrid organic-inorganic trihalide perovskite, namely methylammonium lead bromide (MAPbBr$_3$), from the half metal ferromagnetic electrode, La$_{0.63}$Sr$_{0.37}$MnO$_3$ (LSMO) under an external magnetic field and used in spin- light emitting diode (spin-LED) **(Figure 13b)**.[48] The ferromagnetic half-metallic nature of LSMO has a spin polarization ability close to 100% at the Fermi level. It has been suggested Sr segregation and the surface layer made up of SrO and SrCO$_3$ play an important role in spin injection properties at the hybrid interfacial region, probably it forms a tunneling barrier at the interface.[49] However, it is also well-known that the spin injection efficiency ($\gamma$) at the semiconductor/ ferromagnetic interface is proportional to their conductivities ($\sigma$) ratio ($\sigma_{SC}/\sigma_{FM}$). The conductivity of the ferromagnetic electrode (FM) is higher orders of magnitude compared to the semiconductor (SC), which in turn makes $\gamma \ll 1$, resulting in loss of spin information at the SC/FM interface. Without any use of an external magnetic field or ferromagnetic contact, a spin light emitting diode (spin-LED) can operate at room temperature using a chiral organic molecule which can be explained by the CISS effect and this is indeed a great achievement in optoelectronic applications.

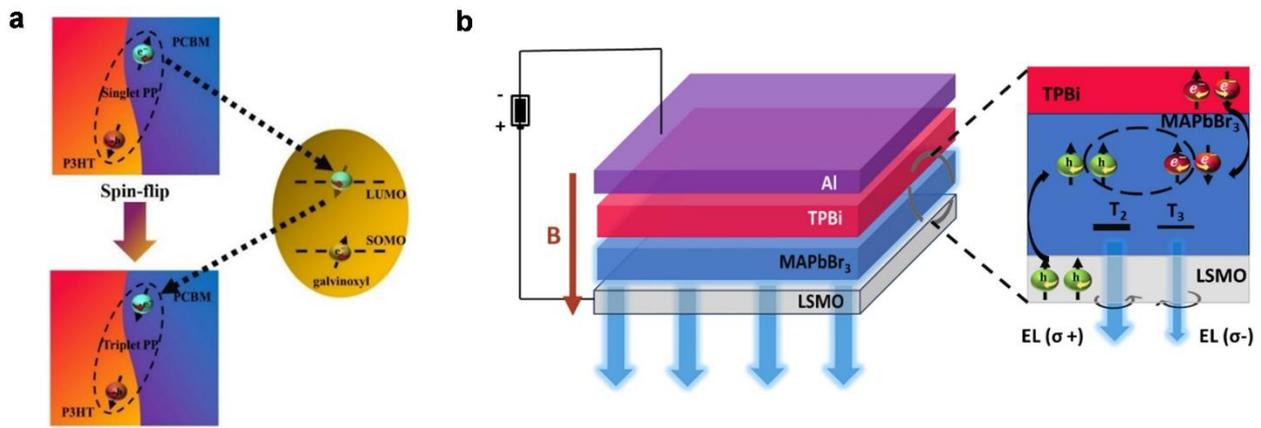

**Figure 13.** (a). Schematic representation of spin-flipping mechanism via galvinoxyl radical. (b). Schematic representation of the spin-LED device architecture consisting of half-metal LSMO anode, MAPbBr$_3$ light emitting layer, and TPBi molecule thin film as the electron transport layer covered with an Al cathode. Figure a is reconstructed from ref.[42] Figure b is adapted with permission from ref.[48]

The effective spin injection from chiral halide perovskite (R/S-MBA)$_2$BX$_4$ (MBA= α-methylbenzylammonium; B = Sn, Pb; X=I, Br) layers into commercially workable III-Vs based semiconductors (e.g., Si, Ge etc.) has been demonstrated. It has been observed that the current injected was spin-polarized, and through the detection of circularly polarized light emission, it has been confirmed the spin accumulation in the III-Vs-based semiconductor. Furthermore, the conductivity mismatch, observed between ferromagnetic electrode and organic semiconductor has been successfully overcome. In addition, the spin orientation formed by the CISS mechanism has paralleled the direction of the current and the emitted light has also been paralleled with the current direction because of the small escape cone of III-Vs-based semiconductor. Thus, it enhances spin-to-light conversion and satisfies optical selection rules that require the circularly polarised light helicity to be parallel with the emission direction. Kim and others have demonstrated control over charge, spin, and light by incorporating the chiral metal-halide perovskite ((R-/S-MBA)$_2$PdI$_4$) hybrid semiconductor that shows the CISS effect **(Figure 14)**.[50] It has been observed that the circularly polarised electroluminescence (CP-EL), which is emitted through the spin-LED has a spin-polarized current of >80% with 2.6% efficiency at room temperature by controlling the intensity of CP-El and proper alignment. They studied spin-polarized charge transport in the CISS layer and CISS/nonchiral nanocrystal (NC) heterostructure by mCP-AFM and observed a higher current for ITO/m-PEDOT: PSS/(R-MBA)$_2$PbI$_4$/CsPbI$_3$ NC film when the tip was magnetized up compared to the tip either magnetized down or nonmagnetized. The reverse observation was for ITO/m-PEDOT: PSS/(S-MBA)$_2$PbI$_4$/CsPbI$_3$ NC film **(Figure 15)**.

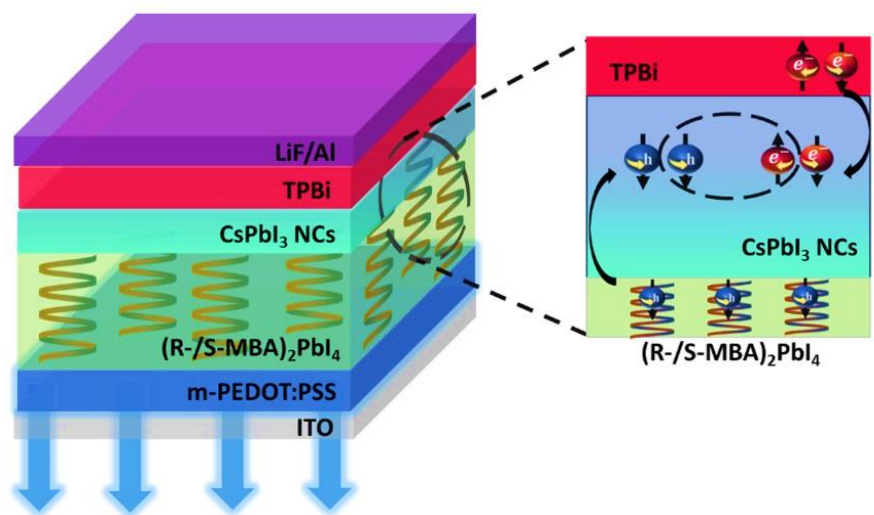

**Figure 14.** Schematic representation, illustrating the spin-polarized charge injection and circularly polarised electroluminescence emission in spin-LEDs as demonstrated in the literature. Figure is reconstructed from ref.[50]

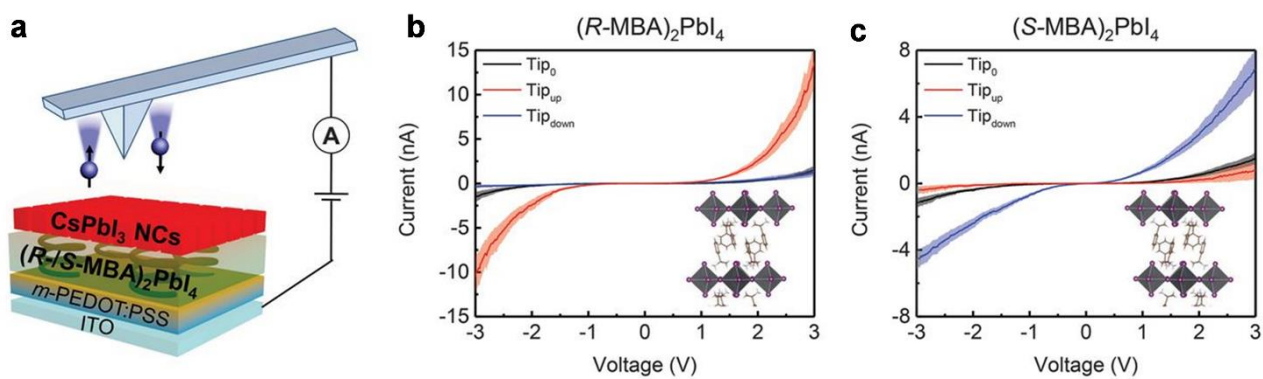

**Figure 15.** (a) Schematic representation of mCP-AFM measurements of the device in the form of ITO/m-PEDOT: PSS/(R-/S-MBA)$_2$PbI$_4$/CsPbI$_3$ NC films. Current vs voltage plots of the device using (b) (R-MBA)$_2$PbI$_4$ (c) (S-MBA)$_2$PbI$_4$ at room temperature. Insets are displaying the crystal structure of (R-/S-MBA)$_2$PbI$_4$. Tip$_0$ represents the nonmagnetized AFM tip; Tip$_{up}$ represents the AFM tip magnetized up; Tip$_{down}$ represents the AFM tip magnetized down. Figures are adapted with permission from ref.[50]

**Oxygen evolution reaction & oxygen reduction reaction**

Organic-inorganic hybrid materials have extensively been used in oxygen evolution reactions as well as oxygen reduction reactions. Specifically, the efficient electrocatalysts, for electrochemical water-splitting, are very crucial in the future for sustainable energy as well as carbon neutrality processes. There is a relation between the activity of an electrocatalyst and its binding energy towards the reaction intermediates. The conventional way to increase the activity is to optimize the binding energy of the catalyst which should be optimal according to the Sabatier principle. However, the sluggish kinetics and high overpotential of OER limits the extensive use of electrocatalytic water-splitting in different aspects of renewable energy applications. Hence, beyond the thermodynamic limitations, that ignore the role of spin, some unconventional approaches have been investigated for the improvement of OER. As the ground state of diatomic oxygen is a triplet, spin constraints should affect the elementary reaction step. By spin-polarized electron transfer, the formation of the reaction intermediate can be controlled and also the efficiency of the OER can be improved. Hence, chiral organic-inorganic hybrid materials have been explored as one of the solutions for the reduction of high overpotential, controlling the formation of the side products and increasing the Faradic efficiency.[26,51–56] Using nonmagnetic or achiral electrodes, the reaction goes through a singlet potential surface, which forms excited state oxygen and thus an energy barrier exists between two energy states. Besides, the chiral or magnetic electrodes, where electrons' spins are coaligned might overcome the issue **(Figure 16)**. Thus, the activity of OER and ORR catalysts can be improved beyond the volcano limit by chiral molecular functionalization. Besides, it offers more versatility with sustainable electron spin polarization and can be implemented into a variety of catalysts, irrespective of their electronic (magnetic) properties under different reaction conditions without affecting the catalyst's composition and surface reconstruction.[17,18,57] In this respect, chiral $Fe_3O_4$ nanoparticles (NPs) have been found to show superior catalytic activity towards oxygen evolution reaction by exhibiting lower overpotential and minimizing $H_2O_2$ production. Significant current density has been obtained in the range of 10 mA/cm$^2$ which is much higher than the previously reported $TiO_2$ coated with a chiral molecules/film monolayer, where the current density was observed below 1 mA/cm$^2$.[58] For chiral modification, $Fe_2O_3$ nanoparticles were encapsulated with chiral amino acids and peptide molecules.[58] The results are shown in **Figure 17** reveal that the current density obtained from the chiral catalyst is much higher than the achiral one. Another work demonstrating the chiral copper oxide films has been employed to improve the chemical selectivity for electrocatalytic water-splitting.[57] The chiral CuO films act as electron spin filters, developed on the electrode surface by electrodeposition method from chiral Cu(II) complex containing electrolyte solution. The work revealed that spin-polarised electrons favored triplet oxygen production, suppressing the $H_2O_2$ byproduct production, low overpotential, and high current density. A similar observation has been found for electrochemically deposited chiral cobalt oxide-coated anode.[59] Vadakkayil *et al.* also demonstrated in detail that doping iron inside the chiral organic molecule modified cobalt oxide decreases the overpotential for OER.[60]

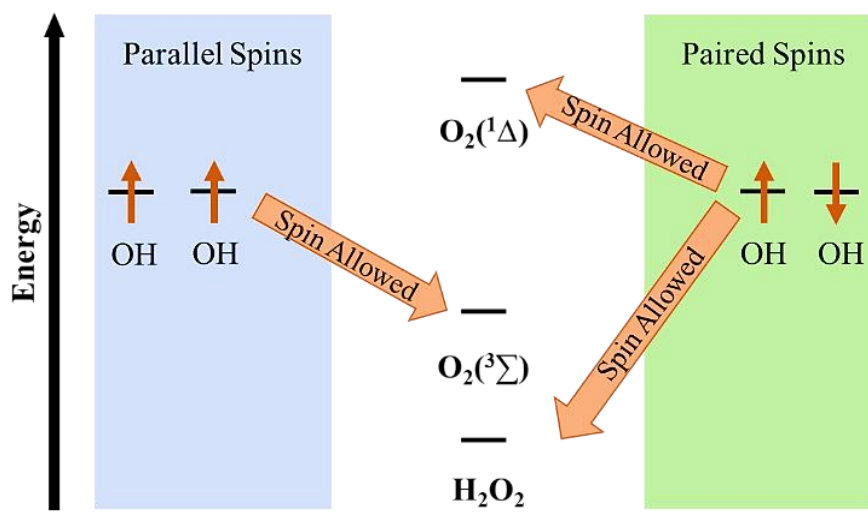

**Figure 16.** Proposed energy diagram as described in the literature.[57] The figure is reconstructed from ref.[57]

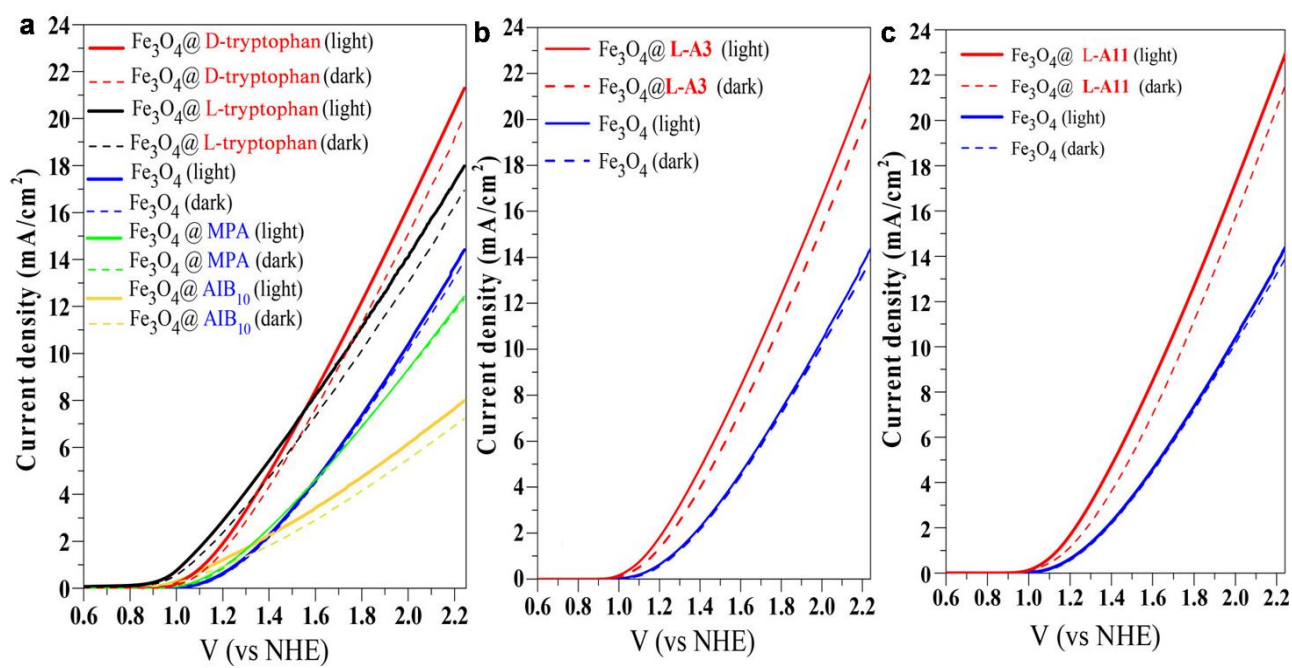

**Figure 17.** (a-c) Linear sweep voltammetry plots of $Fe_2O_3$ nanoparticles encapsulated with different chiral and achiral molecules. Electrochemical water splitting was carried out using 0.1 M KOH solution (pH 13) as the electrolyte and a saturated calomel electrode as a reference electrode, with a scanning rate of 20 mV/s. For chiral modification, L-and D-tryptophan, L-A3 (SH–$(CH_2)_2$–NH–(Ala-AIB)$_3$-COOH, chiral), L-A11(SH-$(CH_2)_2$–NH–(Ala-AIB)$_{11}$–$NH_2$, chiral) have been used whereas as achiral modification 3-mercaptopropionic acid, AIB10 (NH–$(CH_2)_2$–SH–(AIB)$_{10}$–$NH_2$, achiral) have been used. 'Ala' denotes alanine and 'AIB' denotes α-aminoisobutyric *acid*. The figure is adapted with permission from ref.[58] Copyright 2018, American Chemical Society.

Recently, another interesting work by Nair *et. al.* demonstrates the synergistic effects of the inherent magnetic property of iron oxide nanoparticles and chiral molecular functionalization on OER activity.[61] In this study, ferrimagnetic (f) and superparamagnetic (s) $Fe_3O_4$ catalysts were synthesized, and enhancement of OER activity was found for both magnetic particles upon applying an external magnetic field due to spin-selectivity. In addition, s-$Fe_3O_4$ was functionalized with chiral organic molecule R- and S- 1,2-diaminopropane dihydrochloride (DPDC) that results in 42% spin-polarization which significantly enhances the OER activity by increasing the current density and lowering the onset potential as depicted in **Figure 18**.

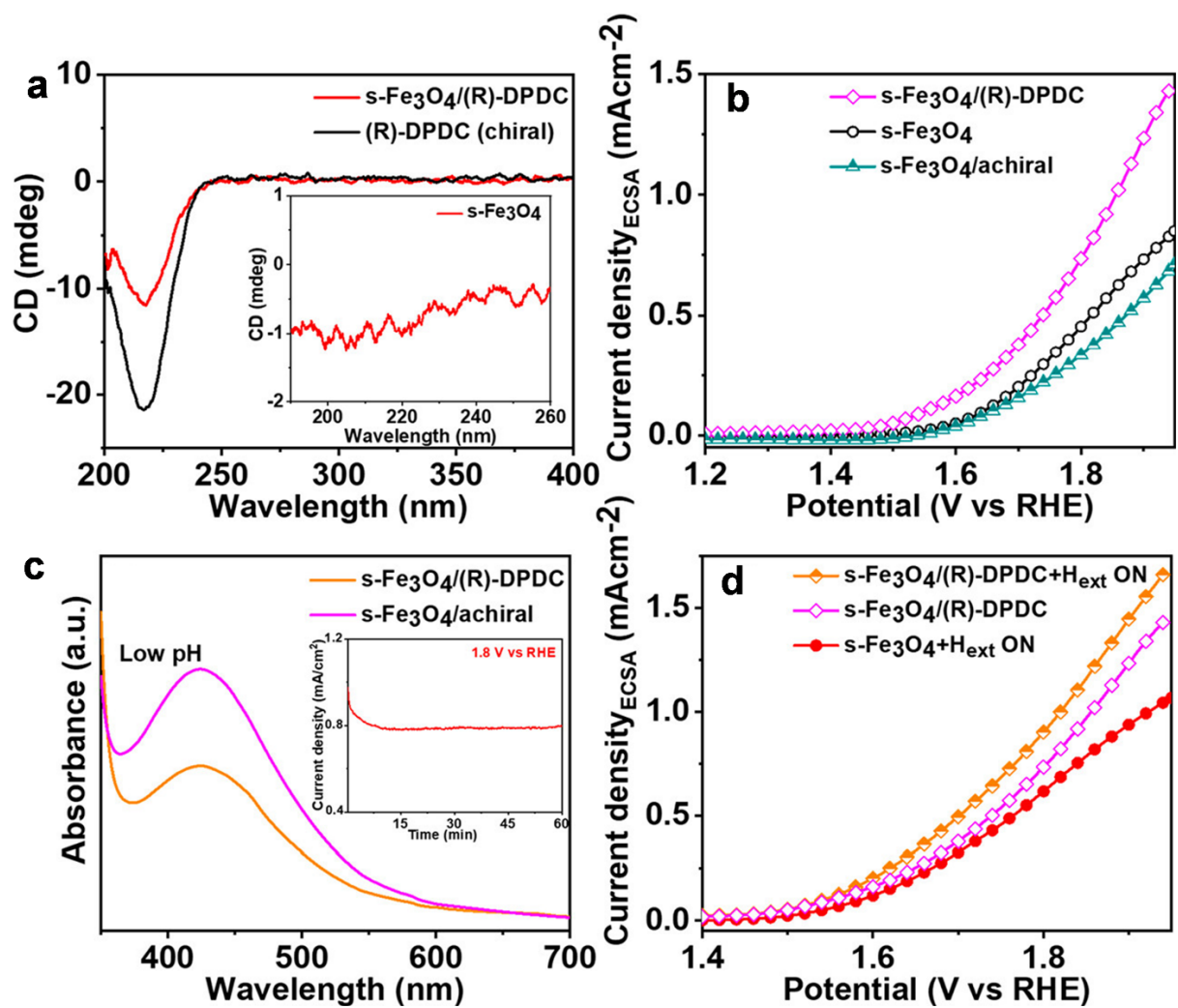

**Figure 18**. (a) Circular dichroism (CD) spectra of chiral DPDC molecules and chiral functionalized superparamagnetic (s) $Fe_3O_4$ catalysts. (b) Linear sweep voltammetry plots exhibiting the augmented electrochemical activity of chiral catalyst. (c) UV−vis absorption spectra for the estimation of $H_2O_2$ detection using chiral and achiral catalysts. (d) LSV curves of the chiral s-$Fe_3O_4$ catalyst under external magnetic field. DPDC is used for chiral modification and ethylene diamine was used for achiral modification. Figures are adapted with permission from ref.[61] Copyright 2023 American Chemical Society.

A recent study shows that chiral modification of iridium nanoparticles also boosts the current density for OER with respect to the pristine metal nanoparticles as well as achiral particles.[62] Feng et al. explored that a chiral amorphous Fe-Ni electrocatalyst synthesized by the electrodeposition method enhances the electrochemical oxygen evolution with respect to its achiral analog by spin-polarized charge transfer process.[63]

In addition to this single metal oxides-based catalyst, two-dimensional transition metal-based oxo hydroxides are also explored as chiral catalysts by integrating them with chiral organic molecules. Recently, Lingenfelder et al. reported that integrating chiral fused thiadiazol-helicene molecules with 2D Ni- and Ni, Fe-based catalyst, spin polarization can be introduced into the anode that helps in augmentation of the OER activity alkaline media.[64] It was demonstrated that Fe doping in the catalysts along with Fe impurities in the electrolytes efficiently enhanced the OER activity **(Figure 19).** The study suggests that the bonding strength and charge transfer between adsorbed oxygen and chiral catalyst are strongly influenced by electron's spin. In this study, the gold surface was functionalized with the chiral helicine molecule and then the NiFe-containing catalyst was coated on the top of the chiral molecule (**Figure 19c**). The superior catalytic activity results from the spin-polarization of the catalyst surface.

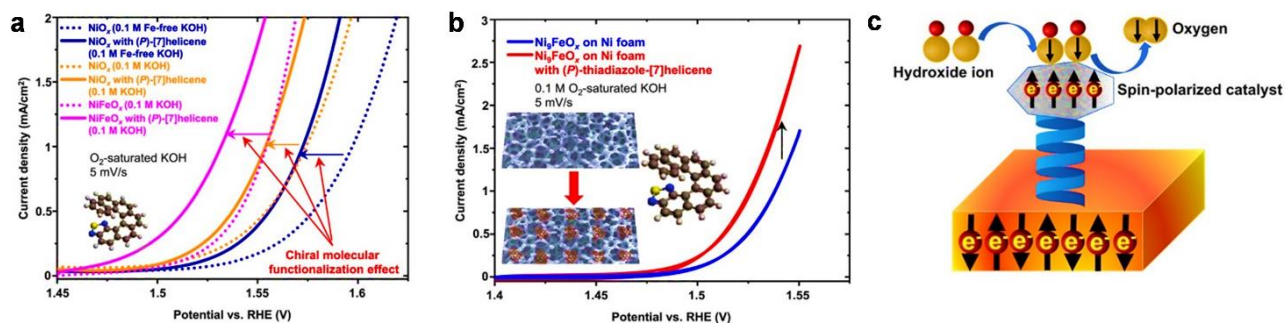

**Figure 19.** Current density vs. applied potential plots (a) to check the effect of chiral modification on $NiO_x$ islands in Fe-free KOH, $NiO_x$ islands in unpurified KOH, and $NiFeO_x$ islands in unpurified KOH. (b) to check the chiral molecular functionalization enhancement at $Ni_xFeO_x$ deposited on Ni foam. (c) Proposed mechanism of OER. Figures are adapted with permission from ref.[64] Copyright 2022, The Author(s).

Apart from OER, chiral hybrid materials are also found to be very effective for the oxygen reduction reaction process. With the growing interest in clean energy technology like fuel cells, and metal-air batteries, the electrocatalyst for oxygen reduction reaction (ORR) is also imperative for sustainability for this type of future clean energy. Like, OER spin selection limitation is also associated with ORR as the diatomic oxygen is involved as substrate here.[65] Sang and others compared two electrodes coated with self-assembled monolayer (SAM) of achiral 3-mercaptopropionic acid (MPA) and chiral

L-cysteine and found an approximately 0.17 V shift in the reduction's onset potential despite the two molecules having the same length and a resembling structure **(Figure 20a)**. They further studied the chirality effect on gold (Au) and platinum (Pt) nanoparticles (NPs). By comparing chiral gold NPs and achiral ones, they found approximately 178 mV lower onset potential for chiral Au NPs than its achiral part **(Figure 20b)** and 80 mV lower onset potential for chiral Pt NPs compared to the achiral Pt NPs and 30 mV lower onset potential compared to commercial Pt/C for ORR process **(Figure 20c)**. The proposed mechanism is displayed in **Figure 20d**.

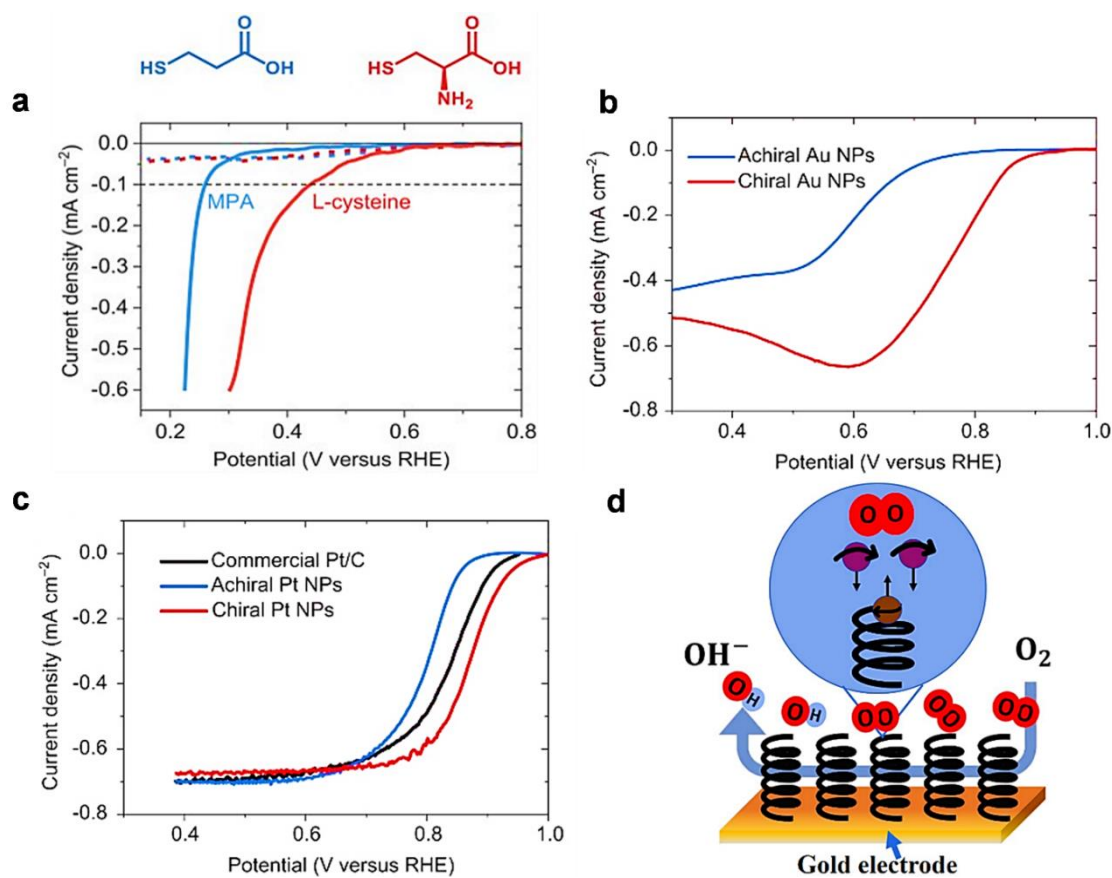

**Figure 20.** (a). LSV plots for the electrode coated with achiral molecules (blue curves) and chiral molecules using (red curves) in $N_2$ (dashed lines) and $O_2$ (solid lines)-saturated 0.1 M KOH (pH 13) aqueous solution as electrolyte. Current density vs. applied potential plots obtained from (b). chiral and achiral Au NPs coated electrodes in oxygen-saturated 0.1 M KOH aqueous solution at a scanning speed of 50 mV/s in static condition (c). chiral and achiral Pt NPs and the commercial catalysts Pt/C coated electrodes in air-saturated 0.1 M KOH solution with a rotation speed of 1,500 rpm and a sweep rate of 5 mV s$^{-1}$ using a rotating disc electrode. Figures a, b, and c are adapted with permission from ref.[65] (d). Schematic representation of chirally modified electrode favoring ORR through a spin-specific way. The scheme is reconstructed from ref.[65] Copyright 2022, PNAS.

Laccase-modified cathode has been shown to combine the two strategies to promote effective electron transfer: (1) "direct electron transfer" and (2) "mediated electron transfer" by the Kashiwagi group enzymatic biofuel cell.[66] Previously it is already stated that some of the electrochemical redox reactions depend on electron spin orientation.[67] The study demonstrates that the chirality of the oligopeptide linker that links the enzyme to the surface influences the ORR efficiency of the laccase-modified electrode. However, Cu(II) complex with amino-acid derivative with/without azobenzene moiety as a mediator enhanced electrochemical current by improving the electron transport between the enzyme and the cathode. Hence, it should be appreciated that the only chiral organic molecule may not compete with the commercially available inorganic catalyst for both OER or ORR but its integration with top-performing catalysts will enhance their efficiency.

**Concluding remarks**

Over the last decade, the CISS effect has experienced substantial growth, with researchers exploring numerous directions and making significant strides. Through experiments, scientists worldwide have observed the presence of different spin polarization in chiral molecules when one contact is magnetic, and the magnetization is switched. The CISS effect manifests in a wide range of temperatures and can be seen in single molecules, monolayers, and thin films. This research shows great promise in providing new insights into various fields, including chemistry, biology, and physics. As fundamental studies continue, they are creating new avenues for CISS research in diverse fields like emergent magnetism, spin-controlled chemistry, and the origin of homochirality in biology. Another crucial area of exploration is the role of CISS in quantum information science (QIS). Developing materials that exhibit quantum properties at room temperature is a significant goal of QIS, and CISS presents a new approach. While the full potential of CISS is yet to be determined, its implications cannot be fully projected at this stage. Therefore, it is vital to continue this research to comprehend its complete potential and pave the way for future discoveries.

**Acknowledgements.** R.G. thanks IIT Kanpur for a senior research fellowship. A.B. acknowledges MHRD for junior research fellowship. A.K.M. acknowledges the DST INSPIRE grant (DST/INSPIRE/04/2021/000055). K.B.G. acknowledges the Start-up Research Grant (SRG/2022/000737) SERB-India. P.C.M. acknowledges the Science and Engineering Research Board, New Delhi (Grant No. CRG/2022/005325).

**Data availability.** No primary research results, software, or code have been included and no new data were generated or analyzed as part of this review article.

**Author contributions.** Conceptualization & supervision: P. C. M.; writing— original draft: R. G., A.B., R. Garg; writing—review and editing: P. C. M., A. K. M. K. B. G.; funding acquisition: P. C. M., A. K. M. K. B. G . All authors have read and agreed to the published version of the manuscript.

**Competing interests.** The authors declare no competing interests related to this work.